\documentclass[a4paper,11pt]{article}
\usepackage{pos}

\usepackage{slashed}
\usepackage{bbold}

\DeclareMathOperator{\Tr}{\rm Tr}

\title{Condensation of lighter-than-physical pions in QCD}
\author[a]{Bastian B. Brandt}
\author*[b]{Volodymyr Chelnokov}
\author[b]{Francesca Cuteri}
\author[a,c]{Gergely Endr\H{o}di}

\affiliation[a]{
  Institute for Theoretical Physics, University of Bielefeld, \\
  D-33615 Bielefeld, Germany
}
\affiliation[b]{
  Institut f\"{u}r Theoretische Physik, Goethe-Universit\"{a}t Frankfurt\\
 Max-von-Laue-Str.\ 1, 60438 Frankfurt am Main, Germany
}
\affiliation[c]{
    Institute of Physics and Astronomy,
ELTE E\"otv\"os Lor\'and University,\\
P\'azm\'any P.\ s\'et\'any 1/A, H-1117 Budapest, Hungary
}

\emailAdd{brandt@physik.uni-bielefeld.de}
\emailAdd{chelnokov@itp.uni-frankfurt.de}
\emailAdd{cuteri@itp.uni-frankfurt.de}
\emailAdd{gergely.endrodi@ttk.elte.hu}

\abstract{We report on the results of the 2+1 flavour QCD simulations at nonzero isospin chemical potential performed at half the physical light quark mass. At low temperatures and large isospin chemical potential Bose-Einstein Condensation (BEC) occurs, creating a pion condensed phase, separated from the hadronic and quark-gluon plasma phases by the BEC transition line. For physical quark masses, the section of this line between the hadronic and BEC phases was found to be almost perfectly vertical, i.e.\ aligned with the temperature axis. We show that for lighter than physical pions, this section remains vertical, and approaches the axis of vanishing chemical potential linearly with the pion mass, giving a prediction of the phase diagram in the chiral limit.}

\FullConference{The 41st International Symposium on Lattice Field Theory (LATTICE2024)\\
 28 July - 3 August 2024\\
Liverpool, UK\\}

\begin{document}
\maketitle

\section{Introduction}

\begin {figure}[b]
\centering
\includegraphics[scale=0.4]{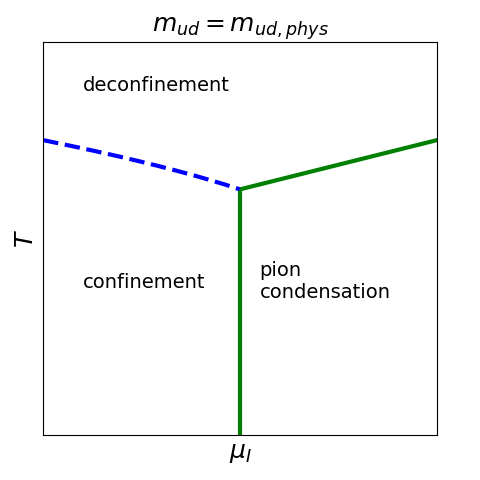}
\includegraphics[scale=0.4]{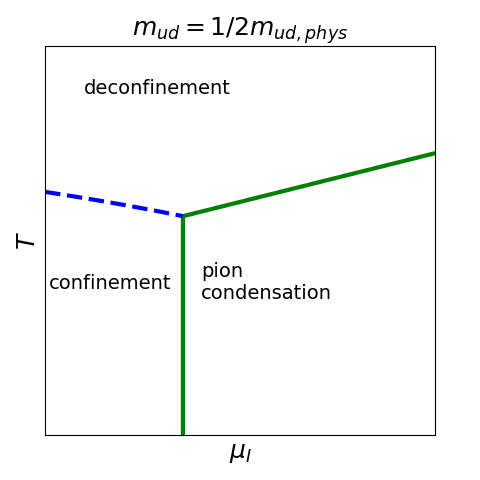}
\includegraphics[scale=0.4]{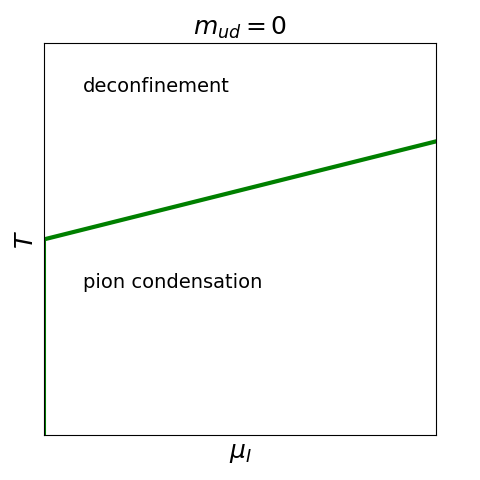}
\caption{A possible scenario for the phase diagram as the chiral limit is approached.}
\label{fig:chiral_scenario}
\end{figure}

While the QCD phase structure at zero matter density is well-studied using Monte-Carlo simulations,
the introduction of a nonzero fermion density (nonzero chemical potential) results in a sign problem 
that prohibits direct sampling in the general case. 
While many approaches to overcome the sign problem and extract the QCD phase structure at nonzero chemical potential exist
(see, for example~\cite{sign-problem-review-Forcrand}), there is currently no conclusive result for large densities. 
As opposed to the finite baryon chemical potential, the theory with only finite isospin chemical potential is sign problem free
and can be simulated directly~\cite{finite-isospin}. 
Apart from being important as an extension of the sign-problem-free QCD region, providing a way to check the algorithms aimed 
at studying the QCD in presence of a sign problem~\cite{taylor-reliability}, 
and providing another starting point to reach the general nonzero fermion density case~\cite{isospin-reweighting},
the theory with nonzero isospin chemical potential might be relevant for the early Universe~\cite{Vovchenko:2020crk}, too.
Some time ago the phase diagram of QCD in $T$-$\mu_I$ plane has been explored~\cite{muI-phys}, including a phase with Bose-Einstein condensation (BEC) of charged pions in addition to the standard hadronic and quark-gluon plasma phases.
In the phase diagram, the chiral crossover line reaches the second order pion condensation line at a pseudo-tricritical point, as shown schematically in the left plot of Fig.~\ref{fig:chiral_scenario}.

Another direction problematic for the numerical simulations is the chiral limit, where the inversion of the fermion operator fails due to the appearance of zero modes.
In the chiral limit, the pion condensation boundary approaches $\mu_I = 0$ at least for $T = 0$. 
If the shape of the pion condensation boundary remains the same as in the physical case, it will approach the chiral limit as shown in 
Fig.~\ref{fig:chiral_scenario} -- in the chiral limit the pion condensation boundary remains on the $\mu_I = 0$ axis up to the chiral transition temperature, with the pion condensate existing at arbitrarily small isospin chemical potential. This scenario is also supported by analytical studies \cite{NJL-chiral-isospin, FRG-chiral-isospin}. If this scenario is realized, this might also affect the nature of the chiral phase transition at zero chemical potential. 
The aim of our work is to check this scenario using direct Monte-Carlo simulations of the theory with lighter than physical quark masses. Preliminary results have already been reported in \cite{LATTICE2022}.
In this proceedings article, we determine the location of the pion condensation boundary at $m_\mathrm{ud} = \frac{1}{2} m_\mathrm{ud, phys}$,
propose an improved method of reweighting in the pion source $\lambda$,
and perform a check of the universality class of the transition using the reweighted results.

\section{Lattice simulation setup} 

Our lattice setup is essentially the same as used in~\cite{muI-phys}, except that the light quark mass set to
a smaller-than-physical value. 
Simulations are performed with 2+1 flavours of stout-improved staggered fermions with two levels of stout smearing and a tree-level Symanzik-improved gauge action. 
The partition function we simulate has the form 
 \begin{equation}
\mathcal{Z} = \int \mathcal{D} U_\mu \, 
    e^{-\beta S_G}\, 
    (\det \mathcal{M}_{\rm ud})^{1/4}\,
    (\det \mathcal{M}_{\rm s})^{1/4} \,,
\label{partition-function}
\end{equation}
where $S_G$ denotes the gauge action and $\mathcal{M}_{\rm ud}$ and $\mathcal{M}_{\rm s}$ 
denote the (combined) light and strange fermion operators, respectively.

\begin{equation}
    \mathcal{M}_{\rm ud} = 
\begin{pmatrix}
  \slashed{D}(\mu_I) + m_{\rm ud} & \lambda \eta_5 \\
 -\lambda \eta_5 & \slashed{D}(-\mu_I) + m_{\rm ud}
\end{pmatrix}\,, \;\;\,
\mathcal{M}_s = \slashed{D}(0) + m_s\,.
\label{quark-operators}
\end{equation} 
\begin{equation}
{\rm det}\;\mathcal{M}_{ud} = {\rm det} \left( \left(\slashed{D}(\mu_I) + m_{ud}\right)^\dagger\left(\slashed{D}(\mu_I) + m_{ud}\right) + \lambda^2 \right) \ .
\label{light-quark-determinant}
\end{equation}

The term $\lambda$ represents a pion source -- an unphysical explicit symmetry breaking term 
defining the direction for the possible spontaneous breaking of the U(1) symmetry at $\mu_I\neq0$. 
The simulations are performed at several nonzero $\lambda$ values and then extrapolated to $\lambda = 0$. 

At the boundary of the BEC phase, the pion condensate 
\begin{equation}
\left\langle \pi^\pm \right\rangle = \frac{T}{V} \frac{\partial \log \mathcal{Z}}{\partial \lambda}
= \frac{T}{2V} \left\langle \Tr \frac{\lambda}{|\slashed{D}(\mu_I)+m_{\rm ud}|^2+\lambda^2} \right\rangle
\label{pion-condensate}
\end{equation}
takes on a non-zero value in the limit $\lambda \to 0$. 
Following \cite{muI-phys}, we define the renormalized pion condensate as 
\begin{equation}
\Sigma_\pi = \frac{m_{ud}}{m_\pi^2 f_\pi^2} \left\langle \pi^\pm \right\rangle \,.
\label{renormalized-pion-condensate}
\end{equation}

The main results are obtained on $24^3 \times 8$ lattices, together with several data points calculated 
on $32^3 \times 10$ and $36^3 \times 12$ lattices to check the effect of finite lattice spacing on the results. The comparison of the 
pion condensate extrapolated to $\lambda=0$ at $\mu_I \approx 0.72\;m_\pi$ and approximately same temperatures,  calculated on different lattices is given in
Table~\ref{table:comparison}. We can see that the effect of the different lattice step $a$ is of the same order as the 
stochastic error estimates.
To obtain the value of $\mu_I$ at the BEC phase boundary for four different temperatures 
between 114 MeV and 142 MeV, we perform a scan in $\mu_I$ with at least five different $\mu_I$ values per temperature.
Furthermore, we also located the BEC phase boundary via a scan in the temperature direction at $\mu_I \approx 0.72 \; m_\pi$, performing simulations at nine different temperatures for this specific isospin chemical potential.
For each set of parameters, three values of the pion source $\lambda$ were simulated to enable an extrapolation to $\lambda=0$.
Each measurement was carried out with at least 200 configurations, separated by 10 updates each, and 1000 thermalization updates. 

\begin{table}[htb]
\centering
\begin{tabular}{c c c c} 
 \hline
 $T$ [MeV] & $\Sigma_\pi$, $N_t=8$ & $\Sigma_\pi$, $N_t=10$ & $\Sigma_\pi$, $N_t=12$ \\ 
 \hline
 114 & 0.763(6)  & 0.781(19) & 0.746(19) \\
 123 & 0.675(12) & 0.680(22) & 0.642(16) \\
 132 & 0.532(26) & 0.585(26) & 0.51(7) \\
 \hline
\end{tabular}
\caption{Comparison of the pion condensate extrapolated to $\lambda=0$ at $\mu_I \approx 0.72\;m_\pi$ obtained from lattices with different $N_t$.}
\label{table:comparison}
\end{table}

\begin {figure}[b]
\centering
\includegraphics[scale=0.4]{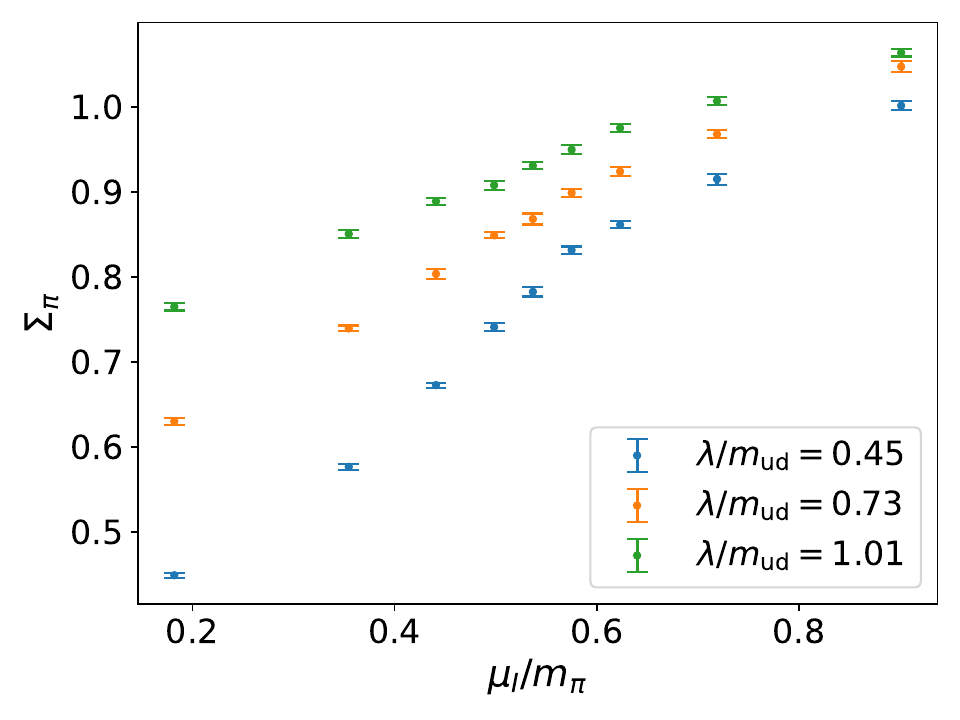}
\caption{Results for the pion condensate vs $\mu_I / m_\pi$ at finite $\lambda$ at $T = 114\; \mathrm{MeV}$ on $N_t=8$ lattices.}
\label{fig:unimproved-pion-condensate}
\end{figure}

The results for the pion condensate obtained using Eq.~(\ref{pion-condensate}) are shown in Fig.~\ref{fig:unimproved-pion-condensate}. 
The figure shows the significant dependence of the condensate on the pion source parameter $\lambda$. 
The extrapolation of this observable to $\lambda = 0$ would require simulations at additional and even smaller $\lambda$ values, in the region where simulations become prohibitively expensive due to larger condition numbers of the Dirac operator. 

\section{Improvements in observable estimation}

To avoid the need to simulate the theory at many small values of $\lambda$, 
we adopt the improvement for the pion condensate with respect to the $\lambda=0$ limit introduced in Ref.~\cite{muI-phys}. The improved observable is obtained by rewriting the trace in Eq.~\eqref{pion-condensate}
as a sum over the singular values of the (massive) Dirac operator $\slashed{D}(\mu_I) + m_{\rm ud}$, which in the infinite volume limit is replaced by an integration over the singular value density $\rho$,
and taking the limit $\lambda \to 0$, resulting in a Banks-Casher type relation for the pion condensate,
\begin{equation}
\lim_{\lambda \to 0} \left\langle \pi^\pm \right\rangle = 
\frac{\pi}{4} \rho(0) \ ,
\label{improved-pion-condensate}
\end{equation}
where $\rho(0)$ is the expectation value of density of the singular values at zero. 

\begin {figure}[tb]
\centering
 \includegraphics[scale=0.3]{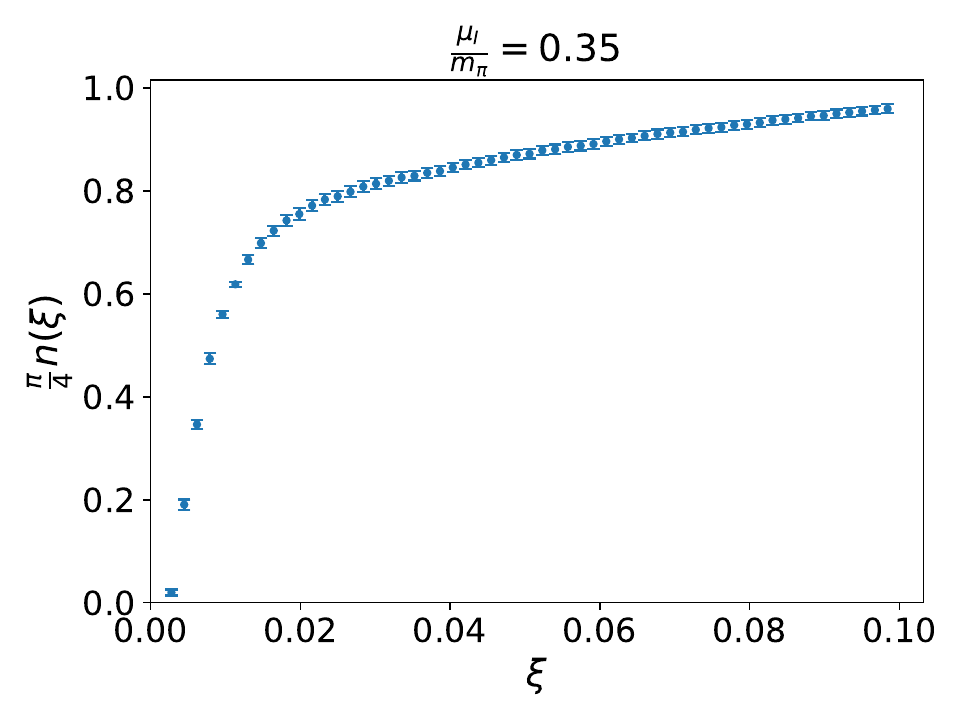}
 \includegraphics[scale=0.3]{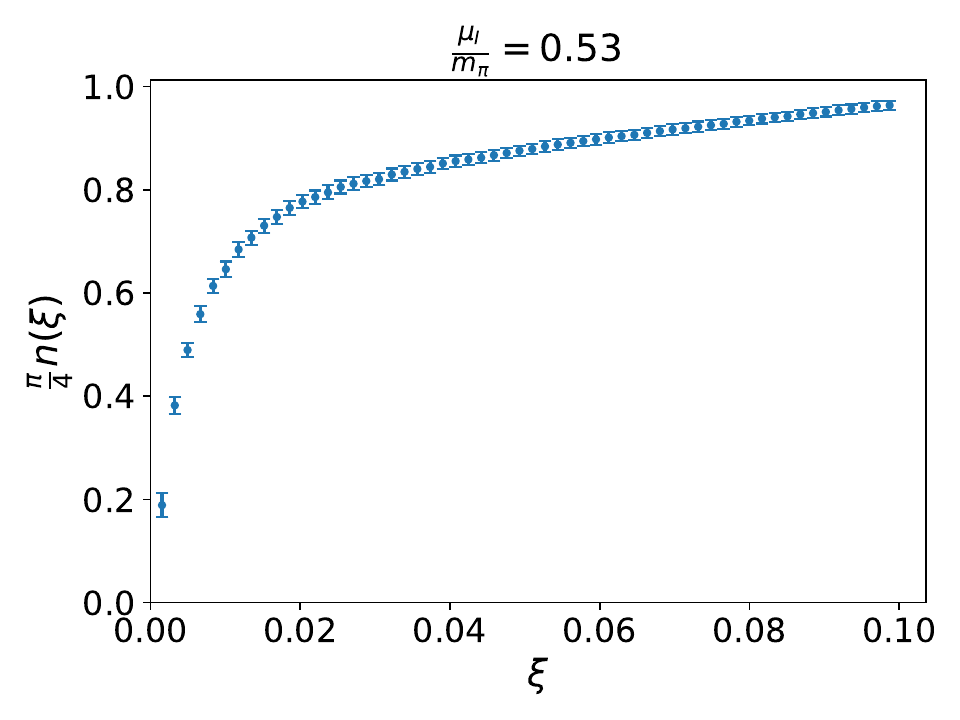} 
 \includegraphics[scale=0.3]{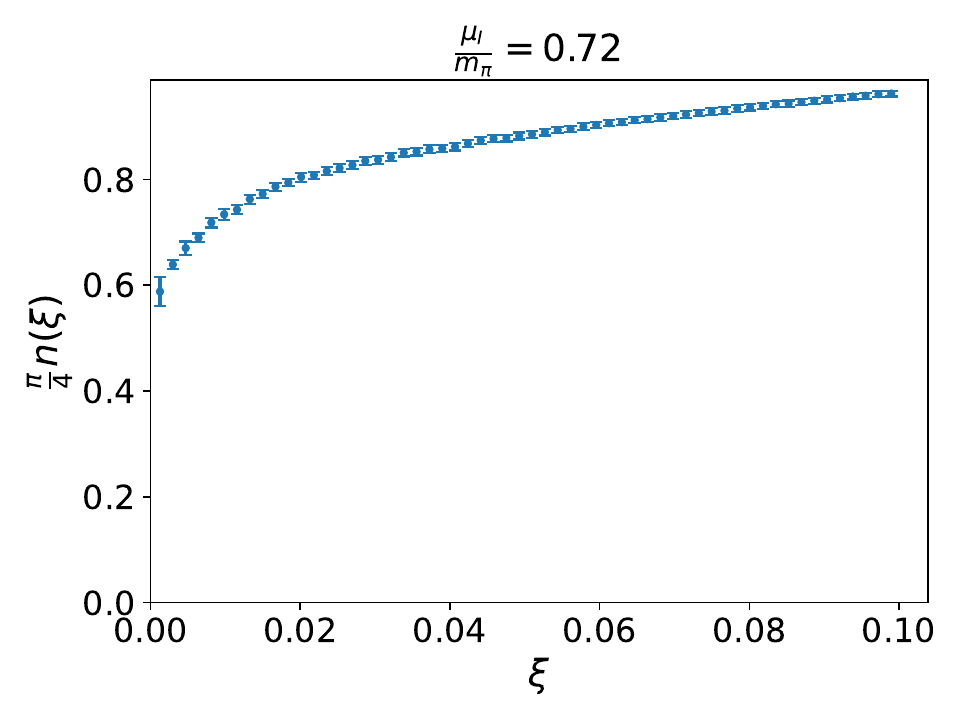} 
\caption{Average singular value density dependence on the window width $\xi$ for the phase with no isospin condensate (left),
the pion condensation boundary (middle), and the pion condensed phase (right).}
\label{fig:singular-value-histograms}
\end{figure}

To extract this value from our simulations, we calculate the 150 smallest singular values of 
the Dirac operator and extrapolate the binned singular value density $n(\xi) \equiv \frac{1}{\xi} \sum_{\xi_i<\xi} 1$
in a given window $[0, \xi]$ to $\xi = 0$,
see also Ref.~\cite{muI-phys}. 
Fig.~\ref{fig:singular-value-histograms} illustrates this process for three different 
values of $\mu_I$ corresponding to the hadronic phase, the pion condensation boundary and to the pion condensed phase.
The phase with no pion condensate (here the hadronic phase) is characterized by the average singular value density going to zero 
already at finite values of the window width $\xi$, leading to a gap between the smallest singular values and zero, while in the BEC phase the average singular value density
goes to a fixed value when $\xi\to0$.

Since the Dirac operator does not have an explicit dependence on $\lambda$, 
the only remaining $\lambda$ dependence for the improved pion condensate comes from the fact that the 
gauge configurations from which the singular values are extracted are sampled with weights corresponding to nonzero $\lambda$ values. 
This dependence is much smaller than the one for the directly measured pion condensate, so one can typically use a linear extrapolation in $\lambda$ to obtain the result at $\lambda = 0$. We can further reduce the $\lambda$ dependence by employing an approximate reweighing in $\lambda$, discussed in the next section. 

\section{Reweighting}

To obtain the correct gauge field distribution at a given $\lambda_\mathrm{new}$ we can make use of a reweighting from the distribution sampled at $\lambda$ 
with weights 
\begin{equation}
W(\lambda_\mathrm{new}, \lambda) = \frac{\left(\det \left[ |\slashed{D}(\mu_I) + m_{ud}|^2 +\lambda_\mathrm{new}^2 \right] \right)^{1/4}}{\left(\det \left[|\slashed{D}(\mu_I) + m_{ud}|^2 + \lambda^2\right]\right)^{1/4}}\ .
\label{exact-reweighing}
\end{equation}
This results in the reweighted observables,
\begin{equation}
\left\langle O \right\rangle_{\lambda_\mathrm{new}} = \frac{\left\langle O\; W(\lambda_\mathrm{new}, \lambda) \right\rangle_\lambda}{\left\langle W(\lambda_\mathrm{new}, \lambda) \right\rangle_\lambda} \,.
\label{single-point-reweighting}
\end{equation}
If the observable $O$ explicitly depends on $\lambda$, then $O$ in right side of Eq.~\eqref{single-point-reweighting} 
is taken at $\lambda = \lambda_\mathrm{new}$ -- the reweighting only corrects for the distribution of the configurations, 
over which the averages are taken.

While calculating the exact determinant ratio is expensive, we can start with a leading-order, i.e.\ linear, approximation~\cite{muI-phys},
\begin{equation}
\log W(\lambda_\mathrm{new}, \lambda) \approx 
- \frac{\lambda^2-\lambda_\mathrm{new}^2}{4} \Tr \frac{1}{|\slashed{D}(\mu_I) + m_{ud}|^2 + \lambda^2} = 
- \frac{\lambda^2-\lambda_\mathrm{new}^2}{\lambda} \frac{V}{ 2 T} \pi^\pm \equiv \log W_{LO}(\lambda_\mathrm{new}, \lambda) \ .
\label{leading-order-reweighing}
\end{equation}

This approximation can be improved using the lowest $k$ singular values via
\begin{equation}
\log W(\lambda_\mathrm{new}, \lambda) = \frac{1}{4} \sum_{i=1}^{N} \log \frac{\xi_i^2 + \lambda_\mathrm{new}^2}{\xi_i^2 + \lambda^2} 
 \approx \log W_{LO}(\lambda_\mathrm{new}, \lambda) + \frac{1}{4} \sum_{i=1}^{k} \left( \log \frac{\xi_i^2+ \lambda_\mathrm{new}^2}{\xi_i^2 + \lambda^2} +  \frac{\lambda^2- \lambda_\mathrm{new}^2}{\xi_i^2 + \lambda^2} \right) \ ,
 \label{beyond-leading-order-reweighing}
\end{equation}
where $N=3N_s^3N_t$ is the total number of singular values on the lattice. The contribution of the remaining singular values is included in the leading order expansion. 

Finally, since we have configurations for three different $\lambda$ values, we can use multihistogram reweighting~\cite{Ferrenberg-Swendsen,Ejiri} to extract the observables at a set of $\lambda_\mathrm{new}$ from the combined data for different $\lambda$, as explained in the following.

Assume that we have $n$ different simulations, the $i$-th of which is done at $\lambda = \lambda_i$ and has $P_i$ (uncorrelated) data points.
Then we can write an expression for the averages of any observable $O$ at $\lambda_\mathrm{new}$, using all collected data, similar to Eq.~ref{single-point-reweighting} using a single simulation at one value of $\lambda$.
We fix an arbitrary pion source value $\lambda_0$, and write $W(\lambda_\mathrm{new}) \equiv W(\lambda_\mathrm{new}, \lambda_0)$. 
Then for a given observable $O$

\begin{equation}
    \left\langle O \right\rangle_{\lambda_\mathrm{new}} = 
    \frac{1}{Z_{\lambda_\mathrm{new}}} 
    \sum_{i = 1}^{n}  
    \sum_{k = 1}^{P_k}
    \frac{O_{i,k} W_{i, k}(\lambda_\mathrm{new})}{\sum_{j=1}^n P_j \; Z_{\lambda_j}^{-1} \; W_{i, k}(\lambda_j)} \ .
    \label{multihistogram-reweighting}
\end{equation}

Here $O_{i, k}$ and $W_{i, k}$ are the values of observable $O$ and weight $W$ on the $k$-th configuration in the $i$-th simulation.
Note that $W_{i, k}(\lambda) = W_{i, k}(\lambda, \lambda_0)$, where $\lambda_0$ is an arbitrary constant that does not change with $i$. 
In particular, this means that the pion condensate observable $\pi^\pm$ in Eqs.~\eqref{leading-order-reweighing} and \eqref{beyond-leading-order-reweighing}
is calculated using $\lambda = \lambda_0$ on every configuration, independently on the $\lambda$ values used to produce that specific configuration. 
In practice, we took the smallest simulated $\lambda$ as $\lambda_0$.
Note also, that the observable $O$ does not explicitly depend on $\lambda$ in our case.
The values of $Z_{\lambda_\mathrm{new}}$ and $Z_{\lambda_j}$, are extracted from the self consistency relation, obtained by setting $O \equiv 1$ in Eq.~\eqref{multihistogram-reweighting}.

The derivation of Eq.~\eqref{multihistogram-reweighting} follows \cite{Ejiri}, with the single difference that the 
weights $W_{i, k}(\lambda)$ have an arbitrary dependence on $\lambda$, so we cannot state $\log W = \lambda U$ for some observable $U$, 
and consider a joint histogram over $U$ and all observables of interest. This can be overcome by either considering a Taylor expansion 
of $\log W(\lambda)$ in $\lambda^2 - \lambda_0^2$ to some high enough order (similarly to what is done in \cite{Ejiri} for the hopping parameter and the chemical potential) or by treating the histogram as the probability distribution of the gauge field configurations, on which the observables are calculated (instead of the distribution of the observables). 
We note that a similar multihistogram reweighting approach was employed recently in~\cite{Endrodi:2025hlb}.

\begin {figure}[tb]
\centering
 \includegraphics[scale=0.45]{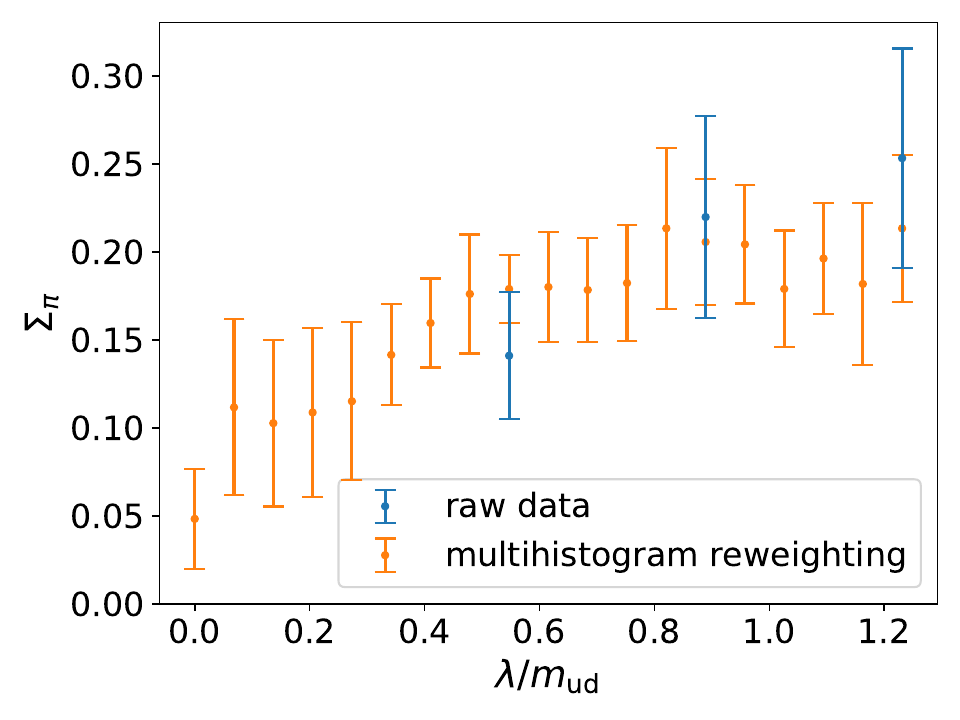}
 \includegraphics[scale=0.45]{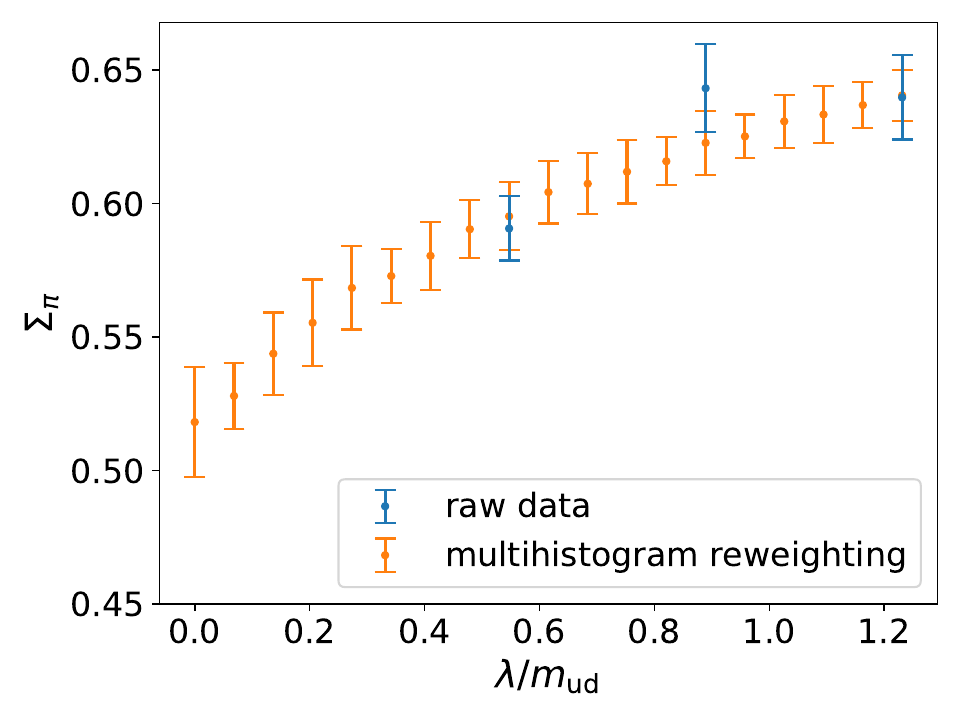} 
\caption{Reweighted improved pion condensate at $T=132\; \mathrm{MeV}$, $\mu_I = 0.53 m_\pi$ (left), $\mu_I = 0.72 m_\pi$ (right).}
\label{fig:multihistogram-reweighing}
\end{figure}

We used the multihistogram reweighting Eq.~\eqref{multihistogram-reweighting} for the singular value densities $n(\xi)$, getting the reweighted singular density distributions and performing the extrapolation to $\xi = 0$ independently for each $\lambda_\mathrm{new}$ to obtain an estimate of $\rho(0)$ at given $\lambda$.
The reweighting results for the improved pion condensate are shown in Fig.~\ref{fig:multihistogram-reweighing}. 
The jagged character of the condensate comes from the need to perform the extrapolation in $\xi$ (shown in Fig.~\ref{fig:singular-value-histograms}),
which is done independently on each reweighted point, thus the errors of the extrapolation, which are responsible for a larger part of the final error are uncorrelated for each sampled $\lambda$ value. 
Note that, while the plots show reweighted values up to $\lambda = 0$, the region where our weight approximation can be trusted does not extend to 0. 
Thus, as a final estimate of the pion condensate at $\lambda = 0$ we use a linear extrapolation of the reweighted pion condensate observables in the region between the smallest and the largest sampled values of $\lambda$ to $\lambda = 0$.

\begin {figure}[tb]
\centering
 \includegraphics[scale=0.6]{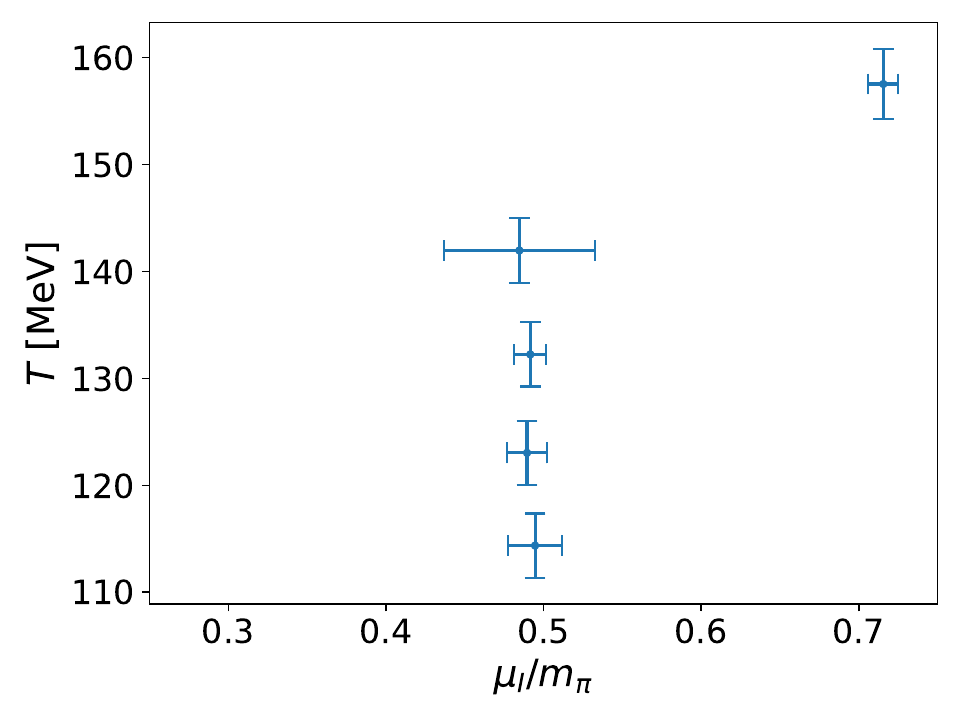}
\caption{Location of the pion condensation line for $m_{\mathrm{ud}} = m_{\mathrm{ud, phys}}/2$.}
\label{fig:condensation-transition}
\end{figure}

After obtaining the pion condensate values at $\lambda = 0$ we use a cubic fit of the data to extract the point where the pion condensate vanishes -- 
the transition point. This was done both for the fixed-temperature scans, resulting in four values of the critical isospin chemical potential for four different temperatures, and for the fixed $\mu_I$ scan, resulting in the critical temperature for $\mu_I \approx 0.72 m_\pi$.
The location of the pion condensation transition extracted from our simulation is shown in Fig.~\ref{fig:condensation-transition}. 
We see that the condensation boundary remains vertical up to $T = 142\; \mathrm{MeV}$, which supports the scenario shown in Fig.~\ref{fig:chiral_scenario}.

We also perform a check of $O(2)$ scaling for the pion condensate in the vicinity of the BEC phase boundary by comparing the data to the form (see \cite{muI-phys} and references therein for the scaling function definition)
\begin{equation}
\label{o2-scaling}
\Sigma_\pi = h^{1/\delta} f_G(t/h^{1/(\beta \delta)}) + a t h + b h \quad \textnormal{with} \quad t = (\mu_c - \mu)/t_0 \quad \textnormal{and} \quad h = \lambda/\lambda_0 \,. \nonumber
\end{equation}
To do that, Eq.~\eqref{multihistogram-reweighting} is applied to extract the $\lambda$  dependence of the unimproved pion condensate (\ref{renormalized-pion-condensate}) for the data points around the pion condensation transition. 
Since $\pi^\pm$ has an explicit dependence on $\lambda$, it needs to be recalculated on all the configurations for every value of $\lambda_\mathrm{new}$.
To avoid that, we used a linear Taylor expansion instead of the exact pion condensate. Similarly to the improved pion condensate reweighting, we had
to limit the smallest $\lambda_\mathrm{new}$ to be around the smallest simulated $\lambda$ -- smaller $\lambda_\mathrm{new}$ values show significant deviation from the scaling, that could be either due to the unreliability of reweighting, or due to insufficient precision of approximation of the pion condensate by the linear Taylor expansion in this region.

\begin {figure}[tb]
\centering
 \includegraphics[scale=0.6]{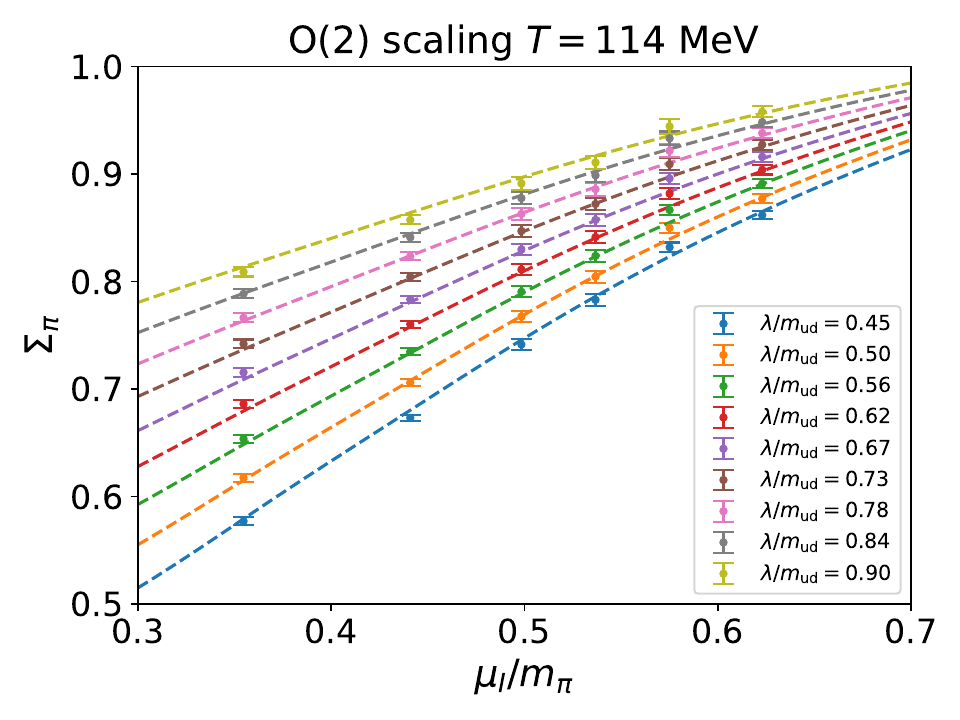}
\caption{Check of the $O(2)$ scaling of the unimproved pion condensate.}
\label{fig:o2-scaling}
\end{figure}

The results are shown in Fig.~\ref{fig:o2-scaling}, which indicates a good description of the unimproved pion condensate in a wide range of $\mu_I$ and $\lambda$ using Eq.~(\ref{o2-scaling})
 ($\chi^2 / \mathrm{dof} = 1.83$, for $0.4 < \lambda / m_\mathrm{ud} < 0.9$, and $0.35 < \mu_I / m_\pi < 0.65$), giving a strong support to the expectation that the BEC phase boundary is a second order transition belonging to the O(2) universality class.

\section{Summary}

In this work we extracted the location of the pion condensation boundary in 2+1 flavour QCD, with the light quark mass equal to half its physical value. The boundary was found to remain vertical up to $T = 142\; \mathrm{MeV}$, supporting the scenario, in which the boundary falls on the $\mu_I = 0$ axis in the chiral limit.
The simulations at smaller than physical quark masses in the pion condensed region at small values of pion source $\lambda$ become numerically expensive due to the ill-conditionedness of the light quark Dirac operator. Using the improved pion condensate observable from the Banks-Casher type relation for the pion condensate
allows us to perform the $\lambda \to 0$ extrapolation from $\lambda \sim m_{ud}$. The extrapolation can be further improved by a multihistogram reweighting in $\lambda$, which we worked out in this contribution. In order to confirm the scenario supported by the results presented above, further simulations at $m_\mathrm{ud} =  m_\mathrm{ud, phys}/4$ are now in progress.

\acknowledgments
This work was supported by the Deutsche Forschungsgemeinschaft (DFG, German Research Foundation) – project number 315477589 – TRR 211. GE also acknowledges funding from the Hungarian National Research, Development and Innovation Office (Research Grant Hungary 150241) and the European Research Council (Consolidator Grant 101125637 CoStaMM). The authors acknowledge the use of the Goethe-NHR and Bielefeld GPU clusters and thank the computing staff for their support.

\end{document}